\documentclass[twocolumn,superscriptaddress,showpacs,nofootinbib,reprint,aps]{revtex4-1}
\pdfoutput=1

\usepackage{graphicx,amsmath,amssymb,bm}

\begin{document}

\title{Efficient calculation 
of chiral three-nucleon forces\\up to N$^3$LO for ab initio studies}

\author{K.\ Hebeler}
\email[E-mail:~]{kaihebeler@physik.tu-darmstadt.de}
\affiliation{Technische Universit\"at Darmstadt, 64289 Darmstadt, Germany}
\affiliation{ExtreMe Matter Institute EMMI, GSI Helmholtzzentrum f\"ur Schwerionenforschung GmbH, 64291 Darmstadt, Germany}

\author{H.\ Krebs}
\email[E-mail:~]{Hermann.Krebs@tp2.ruhr-uni-bochum.de}
\affiliation{Institut f\"ur Theoretische Physik II, Ruhr-Universit\"at Bochum, D-44780 Bochum, Germany}

\author{E.\ Epelbaum}
\email[E-mail:~]{Evgeny.Epelbaum@ruhr-uni-bochum.de}
\affiliation{Institut f\"ur Theoretische Physik II, Ruhr-Universit\"at Bochum, D-44780 Bochum, Germany}

\author{J.\ Golak}
\email[E-mail:~]{ufgolak@cyf-kr.edu.pl}
\affiliation{M. Smoluchowski Institute of Physics,
Jagiellonian University, PL-30348 Krakow, Poland}

\author{R.\ Skibi\'nski}
\email[E-mail:~]{roman.skibinski@uj.edu.pl}
\affiliation{M. Smoluchowski Institute of Physics,
Jagiellonian University, PL-30348 Krakow, Poland}

\begin{abstract} We present a novel framework to decompose three-nucleon forces
in a momentum space  partial-wave basis. The new approach is computationally
much more efficient than previous methods and opens the way to ab initio studies
of few-nucleon scattering processes, nuclei and nuclear matter based on 
higher-order chiral 3N forces. We use the new framework to calculate matrix elements
of chiral three-nucleon forces at N$^2$LO and N$^3$LO in large basis spaces and
carry out benchmark calculations for neutron matter and symmetric nuclear
matter. We also study the size of the individual three-nucleon force
contributions for $^3$H. For nonlocal regulators, we find that
the sub-leading terms, which have been neglected in most calculations so far,
provide important contributions. All matrix elements are calculated and stored
in a user-friendly way, such that values of low-energy constants as well as the
form of regulator functions can be chosen freely.
\end{abstract}

 
\maketitle

\section{Introduction}

The importance of chiral three-nucleon (3N) forces has been demonstrated in
numerous microscopic calculations  of few-body scattering processes, nuclei and
nuclear matter, see~\cite{KalantarNayestanaki:2011wz,3Nrev} for recent review
articles. Contributions to 3N forces (3NFs) start to appear at 
next-to-next-to-leading-order (N$^2$LO) in the chiral expansion within Weinberg's power
counting framework~\cite{Weinberg_EFT,vanKolck:1994yi,Epelbaum_RevModPhys,Machleidt:2011zz},
whereas the subleading chiral 3NFs at next-to-next-to-next-to-leading-order (N$^3$LO)
have the particular feature that they do not include any new low-energy
constants~\cite{Ishikawa:2007zz,Bernard_N3LO1,Bernard_N3LO2}. So far, all 
microscopic calculations for finite nuclei of mass $A > 3$ based on chiral EFT
interactions were limited to 3NFs at leading order (N$^2$LO), whereas NN forces
are commonly included up to N$^3$LO, see \cite{N4LO_NN} for a fifth
order analysis of NN scattering. For NN forces, it is known that N$^3$LO
contributions are important for a high-precision fit of scattering phase shifts
and mixing angles up to laboratory energies of $E \sim
200\,$MeV~\cite{EM,Epelbaum:2004fk,Ekstroem,Epelbaum_RevModPhys,localNN_QMC,improvedlocalNN}. It is clearly
desirable to extend these studies and perform \emph{complete} 
N$^3$LO 
calculations, which would require the inclusion of subleading N$^3$LO 3N
contributions, see Refs.~\cite{GolakN3LO,fullN3LO,Kruger:2013} for first
calculations along this line. In fact, such studies are a central goal of the
recently formed Low Energy Nuclear Physics International Collaboration
(LENPIC)~\cite{LENPIC_web}.

\begin{figure*}[t]
\includegraphics[scale=0.82]{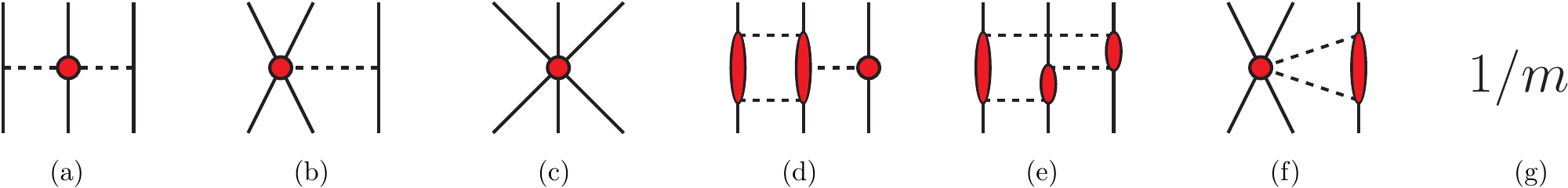}
\caption{(color online) Different topologies that contribute to the
  chiral 3NF up to N$^3$LO (and N$^4$LO). Nucleons and pions are
  represented by solid and dashed lines, respectively.  The shaded
  vertices denote the amplitudes of the corresponding interaction.
  Specifically, the individual diagrams are: (a) 2$\pi$ exchange, 
  (b) 1$\pi$-contact, (c) pure contact, (d) 2$\pi$-1$\pi$ exchange, 
  (e) ring contributions, (f) 2$\pi$-contact and (g) relativistic 
  corrections. See main text for details.}
  \label{fig:topologies}
\end{figure*}

In particular, one may expect that the unresolved
discrepancies for spin-dependent nucleon-deuteron (Nd) scattering observables between calculations
based on high-precision phenomenological NN and 3NF models and experimental
data at intermediate and higher energies cannot be probed at the N$^2$LO level
of accuracy. Furthermore, recent advances in nuclear structure theory make it possible
to extend the range of ab initio calculations to mass numbers of $A =
132$~\cite{heavynuclei}. The inclusion of 3NF contributions beyond the
leading ones in
few- and many-body calculations is the key for increasing the accuracy of
predictions for nuclear observables and for systematic investigations of the
role of chiral symmetry in nuclei and nuclear matter. We also note
that corrections to the 3NF beyond N$^2$LO  (see Fig.~\ref{fig:topologies}) contain such contributions, which 
have already been found earlier in phenomenological interaction models to be
important for the accurate description of properties of light nuclei~\cite{Pieper_3NF}.

Numerous N$^2$LO calculations of Nd scattering at low energy
have revealed, generally, a good agreement between theory and experimental data,
see~\cite{Epelbaum_RevModPhys,Progpart_Epelbaum} and references therein. On the
other hand, the well-known discrepancies such as, e.g., the $A_y$ puzzle and the
cross section in the symmetric space star configuration of the deuteron break up
process could not be resolved at this chiral order~\cite{KalantarNayestanaki:2011wz}. 
Promising results based on chiral 3NFs at N$^2$LO were reported by various groups in
nuclear structure calculations showing, in particular, sensitivity to the
individual 3NF contributions, see Refs.~\cite{3Nrev,Holt:2013fwa} for review
articles. Also recent lattice simulations and no-core shell model calculations 
of light nuclei within chiral effective field theory demonstrate clearly
the important role of the N$^2$LO 3NFs 
(see e.g., Refs.~\cite{Epelbaum:2009pd,Epelbaum:2011md,Epelbaum:2012qn,Roth_NCSM,Barrett:2013nh}).
Furthermore, for neutron-rich systems, contributions from 3NFs have been shown to be of critical
importance for the shell structure and stability close to the neutron drip
line~(see, e.g.,~\cite{calciumnature,oxygenOtsuka,Holt:2010yb,CC,SCGF,CCOakRidge}), while 
for nuclear matter, saturation was demonstrated to be driven by 3NFs~\cite{Bogner:2005sn,HebelerSNM}.

While the formal expressions for the individual contributions of all the
topologies shown in Fig.~\ref{fig:topologies} have already been worked out~\cite{Bernard_N3LO1,Bernard_N3LO2},
their decomposition in a suitable form for few- and many-body frameworks
represents a highly nontrivial task~\cite{Navratil3N,Golak_aPWD,Skibinski_aPWD}.
Due to the huge amount of computational resources needed for this decomposition,
matrix elements have been so far available only in a limited model
space~\cite{GolakN3LO}. As a consequence, consistent N$^3$LO three-body
scattering calculations were limited to low energies and no studies of heavier
nuclei were possible. In this paper we present a novel framework that allows
one to decompose 3N interactions in a plane-wave partial wave basis in a
computationally much more efficient way than the framework of
Refs.~\cite{Golak_aPWD,Skibinski_aPWD}. This new method makes explicit use of
the fact that all (unregularized) contributions to chiral 3NFs are either
local, i.e. they depend only on momentum transfers, or they contain only
polynomial non-local terms. 

In Section~\ref{sec:PWD_local} we derive the new framework
for decomposing local 3NFs efficiently in a momentum-space partial wave
basis. In Section~\ref{sec:app_3NF} we apply the calculated matrix elements of chiral
3NFs up to N$^3$LO to nuclear matter and $^3$H, discuss the partial wave convergence and 
investigate the importance of the individual topologies at different orders in the 
chiral expansion. In Section~\ref{sec:summary} we summarize and given an outlook
of future applications.

\section{Partial wave decomposition of local three-nucleon forces}
\label{sec:PWD_local}

A general translationally invariant 3NF can be expressed as a function of the
Jacobi momenta $\mathbf{p} = \frac{\mathbf{k}_1 - \mathbf{k}_2}{2}$ and
$\mathbf{q} = \frac{2}{3} \left[ \mathbf{k}_3 - \frac{1}{2} (\mathbf{k}_1 +
\mathbf{k}_2) \right]$, where $\mathbf{k}_i$ denote the single nucleon momenta
(in the following equations we will first suppress spin and isospin degrees of freedom):
\begin{equation}
V_{123} = V_{123} (\mathbf{p},\mathbf{q},\mathbf{p}',\mathbf{q}').
\end{equation}
Here and in the following $\mathbf{p}$ and $\mathbf{q}$ ($\mathbf{p}'$ and $\mathbf{q}'$) denote the Jacobi
momenta of the initial (final) state. For local interactions, however, the momentum dependence further simplifies 
as such forces only depend on momentum transfers, i.e. on differences of Jacobi momenta:
\begin{eqnarray}
V_{123}^{\rm{loc}} = V_{123}^{\rm{loc}} (\mathbf{p}'-\mathbf{p}, \mathbf{q}' - \mathbf{q}) \equiv V_{123}^{\rm{loc}} (\tilde{\mathbf{p}}, \tilde{\mathbf{q}}). 
\end{eqnarray}
Note that this statement refers to unregularized forces. Below we will apply non-local 
regulators to the partial-wave decomposed matrix elements. The regularization will be 
discussed in more detail in Section~\ref{sec:app_3NF}.

Generally, the decomposition of 3NFs in plane-wave partial waves involves the evaluation of projection integrals of the form
\begin{eqnarray}
&& F_{L l L' l'}^{m_L m_l m_{L'} m_{l'}} (p, q, p', q') = \int d\hat{\mathbf{p}}' d\hat{\mathbf{q}}' d\hat{\mathbf{p}} d \hat{\mathbf{q}} \nonumber \\
&& \hspace{0.2cm} \times Y_{L' m_{L'}}^{*} (\hat{\mathbf{p}}') Y_{l' m_{l'}}^{*} (\hat{\mathbf{q}}') Y_{L m_L} (\hat{\mathbf{p}}) Y_{l m_l} (\hat{\mathbf{q}}) V_{123}^{\rm{loc}} (\tilde{\mathbf{p}},
\tilde{\mathbf{q}}) 
\label{eq:F_func}
\end{eqnarray}
for fixed values of $p=|\mathbf{p}|, q = |\mathbf{q}|, p' = |\mathbf{p}'|, q' = |\mathbf{q}'|$ and the angular momentum quantum numbers. By using symmetries, 
it is possible to reduce the dimensionality of the angular integrals from 8 to 5. Traditional methods are based on a direct discretization 
and numerical evaluation of these angular integrals~\cite{Golak_aPWD,Skibinski_aPWD}. Due to the large number of external quantum numbers and momentum mesh points 
such algorithms require very large computational resources for calculating all matrix elements necessary for many-body studies. As a 
result, the number of matrix elements of chiral N$^3$LO interactions were so far insufficient for studies of nuclei and matter. 
However, it is possible to evaluate the basic function $F$ defined in Eq.~(\ref{eq:F_func}) in a much more efficient way by explicitly making use 
of the local nature of the 3NFs. Indeed, using rotation invariance of the potential $V_{123}^{{\rm loc}}$ we can write it as a function of three independent variables:
\begin{equation}
V_{123}^{\rm{loc}} (\tilde{\mathbf{p}}, \tilde{\mathbf{q}})=V_{123}^{loc} (\tilde{p}, \tilde{q}, \cos \theta_{\tilde{\mathbf{p}} \tilde{\mathbf{q}}}),
\end{equation}
where
\begin{equation}
\cos \theta_{\tilde{\mathbf{p}} \tilde{\mathbf{q}}}=\frac{\tilde{\mathbf{p}}\cdot \tilde{\mathbf{q}}}{\tilde{p} \tilde{q}}, \quad \tilde{p} = |\tilde{\mathbf{p}}|, \quad\tilde{q} = |\tilde{\mathbf{q}}|. 
\end{equation}
This already shows that the original eight-dimensional integral
contains actually only three non-trivial integrations. The other five integrations, after employing some 
integral transformations, which are described in the appendix, 
can be performed analytically. The final result is given by
\begin{eqnarray}
&& \hspace{-1cm} F_{L l L' l'}^{m_L m_l m_{L'} m_{l'}} (p, q, p', q') \nonumber \\
&=&  \delta_{m_L - m_{L'}, m_{l'} - m_l} (-1)^{m_L + m_{l'}} \frac{2 (2 \pi)^4}{p p' q q'} \nonumber \\
&&\times \sum_{\bar{l} = \text{max} (|L'-L|,|l' - l|)}^{\text{min} (L' + L, l' + l)} 
\frac{\mathcal{C}_{L' - m_{L'} L m_L}^{\bar{l} -m_{L'} + m_L} \mathcal{C}_{l' -m_{l'} l m_l}^{\bar{l} -m_{l'} + m_l}}{2 \bar{l} + 1} \nonumber \\
&&\times \int_{|p'-p|}^{p'+p} d \tilde{p} \, \tilde{p} \int_{|q'-q|}^{q'+q} d \tilde{q} \, \tilde{q} \nonumber \\
&&\times \left. \mathcal{Y}_{L' L}^{\bar{l} 0} (\widehat{\tilde{p} \mathbf{e}_z + \mathbf{p}},\hat{\mathbf{p}}) \right|_{\phi_p = 0, \hat{p} \cdot \mathbf{e}_z = \frac{p'^2 - p^2 - \tilde{p}^2}{2 \tilde{p} p}} \nonumber \\
&&\times \left. \mathcal{Y}_{l' l}^{\bar{l} 0} (\widehat{\tilde{q} \mathbf{e}_z + \mathbf{q}}, \hat{\mathbf{q}}) \right|_{\phi_q = 0, \hat{q} \cdot \mathbf{e}_z = \frac{q'^2 - q^2 - \tilde{q}^2}{2 \tilde{q} q}} \nonumber \\
&&\times \int_{-1}^1 d \cos \theta_{\tilde{\mathbf{p}} \tilde{\mathbf{q}}} P_{\bar{l}} (\cos \theta_{\tilde{\mathbf{p}} \tilde{\mathbf{q}}}) V_{123}^{\rm{loc}} (\tilde{q},\tilde{p},\cos \theta_{\tilde{\mathbf{p}} \tilde{\mathbf{q}}})
\label{final_result_pwd}
\end{eqnarray}
with
\begin{equation}
\mathcal{Y}_{l_a l_b}^{L M} (\hat{\mathbf{a}}, \hat{\mathbf{b}}) = \sum_{m_a, m_b} \mathcal{C}_{l_a m_a l_b m_b}^{L M} Y_{l_a}^{m_a} (\hat{\mathbf{a}}) Y_{l_b}^{m_b} (\hat{\mathbf{b}}).
\end{equation}
Realistic nuclear forces also depend on the spin and isospin quantum numbers of the nucleons. As a complete basis, we choose the standard $Jj$-coupled three-body plane-wave basis of the form~\cite{Gloeckle_book}
\begin{equation}
\left| p q \alpha \right> \hspace{-1.6mm} \phantom \rangle \equiv \left| p q; \left[ (L S) J (l s) j \right] \mathcal{J} (T t) 
\mathcal{T} \right> \, ,  \label{eq:Jj_bas}
\end{equation}
where $L$, $S$, $J$ and $T$ denote the relative orbital angular momentum, spin,
total angular momentum and isospin  of particles 1 and 2 with relative momentum
$p$. The quantum numbers $l$, $s=1/2$, $j$ and $t=1/2$ label the  orbital
angular momentum, spin, total angular momentum and isospin of particle $3$
relative to the  center-of-mass of the pair with relative momentum $p$. The
quantum numbers $\mathcal{J}$ and $\mathcal{T}$  define the total three-body
angular momentum and isospin (for details see Ref.~\cite{Gloeckle_book}). In
Eq.~(\ref{eq:Jj_bas}) we have used the fact that the 3NFs do not depend on the
projections  $m_{\mathcal{J}}$ and $m_{\mathcal{T}}$, hence we omit these
quantum numbers in the basis states here and in the following. The detailed
basis sizes for the different three-body channels for our calculations of
N$^3$LO  matrix elements are presented in Table~\ref{tab:PW_data}. Notice that
as will be shown below, it  is sufficient for our purposes to truncate the value
of the total three-body angular momentum at $\mathcal{J} =9/2$. We have also
verified that further increasing the values of $J_{\rm max}$ leads to negligibly
small effects for all performed calculations. However, for future studies of
heavier nuclei we could extend the calculations to even larger $\mathcal{J}$ and
$J_{\rm max}$, if necessary.

\begin{table}[t]
\begin{center}
\caption{Dimension and file size of the individual 3NF matrix element files up
to N$^3$LO for the different three-body partial waves.  All matrix elements are
calculated and stored in such a way that values of the low-energy couplings
$c_1, c_3, c_4, c_D, c_E, C_S$ and $C_T$ can be chosen freely for the different
topologies, leading to in total 12 files for each partial wave (see main text).
For all partial waves $N_p = N_q = 15$ has been used. $N_{\alpha}$ denotes the
number of partial-wave channels defined in Eq.~(\ref{eq:Jj_bas}). All given
values apply to both three-body parities.}
\begin{tabular}{ccccc}
\hline
$\mathcal{J}$ & $\mathcal{T}$ & $J_{\rm{max}}$ & $N_{\alpha}$ & filesize [GB] \\
\hline
1/2 & 1/2 & 8 & 66 & 0.8 \\
3/2 & 1/2 & 8 & 126 & 3.0 \\
5/2 & 1/2 & 8 & 178 & 6.0 \\
7/2 & 1/2 & 7 & 190 & 6.8 \\
9/2 & 1/2 & 6 & 178 & 6.0 \\
1/2 & 3/2 & 8 & 34 & 0.2 \\
3/2 & 3/2 & 8 & 65 & 0.8 \\
5/2 & 3/2 & 8 & 92 & 1.6 \\
7/2 & 3/2 & 7 & 91 & 1.6 \\
9/2 & 3/2 & 6 & 94 & 1.7 \\
\hline
\label{tab:PW_data}
\end{tabular}
\end{center}
\end{table}

Due to the explicit momentum dependence of the spin-momentum operators, the
knowledge of the function $F$ in Eq.~(\ref{eq:F_func}) is, in general, not
sufficient and the framework needs to be extended.  This can be done in a
straightforward way by factorizing out the momentum dependence of spin-momentum
operators in the form
\begin{equation}
\boldsymbol{\sigma} \cdot \mathbf{x} = \sqrt{\frac{4 \pi}{3}} x \sum_{\mu=-1}^1 Y^*_{1 \mu} (\hat{\mathbf{x}}) \, \boldsymbol{\sigma} \cdot \mathbf{e}_{\mu} \label{eq:spin_factorize}
\end{equation}
and combining the additional spherical harmonic function with the ones in Eq.~(\ref{eq:F_func}) by using
\begin{eqnarray}
&& Y_{l m} (\hat{\mathbf{x}}) Y_{1 \mu} (\hat{\mathbf{x}}) = \nonumber \\
&& \quad \sum_{\bar{L} = |l-1|}^{l+1} \sqrt{\frac{3}{4 \pi} \frac{2 l + 1}{2 \bar{L}+1}} \mathcal{C}_{l 0 1 0}^{\bar{L} 0} \mathcal{C}_{l m 1 \mu}^{\bar{L} m + \mu} Y_{\bar{L} m + \mu} (\hat{\mathbf{x}}), \label{eq:Y_identity}
\end{eqnarray}
where $\mathbf{x}$ represents a Jacobi momentum. This strategy makes it possible
to reduce the expressions for arbitrary spin-dependent interactions to the
expression for spin-independent interactions times some momentum-independent
spin operators. This step has to be performed for each momentum vector in the
spin-momentum operators. Obviously, the efficiency of the present algorithm
decreases with each additional sum over the quantum numbers $\mu$ and $\bar{L}$
in  Eqs.~(\ref{eq:spin_factorize}) and (\ref{eq:Y_identity}). Note, however,
that each of these sums contains only three terms at most.

In order to factorize the momentum, spin and isospin space, for practical calculations 
we first calculate the interaction matrix elements in a $LS$-coupled basis:
\begin{equation}
\left| p q \beta \right> \hspace{-1.6mm} \phantom \rangle \equiv \left| p q; \left[ (L l) \mathcal{L} (S s) \mathcal{S} \right] \mathcal{J} (T t)
\mathcal{T} \right> \, ,\label{eq:LS_bas}
\end{equation}
and recouple only at the end to the $Jj$-coupled basis defined in
Eq.~(\ref{eq:Jj_bas}). Each time the factorization in
Eq.~(\ref{eq:spin_factorize}) is applied, the spin matrix  element effectively
becomes dependent on the quantum number $\mu$, i.e. generally the matrix element
in spin space can be written in the form
\begin{equation}
\left< (S s) \mathcal{S} m_{\mathcal{S}} | \hat{O}_{\sigma} ( \{ \mu_i \} ) | (S' s') \mathcal{S}' m_{\mathcal{S}'} \right>, \label{eq:spin_operator} 
\end{equation}
where the index $i$ counts the number of momentum vectors in the spin operator.
In the same way the function $F$ in Eq.~(\ref{eq:F_func}) becomes a function of
the  quantum numbers $\mu_i$, i.e. it takes the form $F_{L l L' l'}^{m_L m_l
m_{L'} m_{l'} \{ \mu_i \}}$. To be explicit, if we consider, e.g., the case
$\mathbf{x} = \mathbf{p}$ in Eq.~(\ref{eq:spin_factorize}), the function $F$
takes the form
\begin{eqnarray}
&& F_{L l L' l'}^{m_L m_l m_{L'} m_{l'} \mu} (p, q, p', q') = p \sum_{\bar{L} = |L-1|}^{L+1} \sqrt{\frac{2 L + 1}{2 \bar{L}+1}} \nonumber \\
&& \qquad \times \mathcal{C}_{L 0 1 0}^{\bar{L} 0} \mathcal{C}_{L m_L 1 \mu}^{\bar{L} m_L + \mu} F_{\bar{L} l L' l'}^{m_L m_l m_{L'} m_{l'}} (p, q, p', q'), \label{eq:F_withonemu}
\end{eqnarray} 
where we included the factor $\sqrt{\frac{4 \pi}{3}}p$ from the spin operator factorization in Eq.~(\ref{eq:spin_factorize}) in this function.

For an efficient calculation it is important to note that all quantities that
depend on the projection quantum numbers $m$ and $\mu$ are momentum independent.
Hence, it is  advantagous to factorize this dependence in the function $F$.
Specifically, for the example shown in Eq.~(\ref{eq:F_withonemu}) we can write
\begin{eqnarray}
&& F_{L l L' l'}^{m_L m_l m_{L'} m_{l'} \mu} (p, q, p', q') \equiv \delta_{m_L - m_{L'}, m_{l'} - m_l} (-1)^{m_L + m_{l'}} \nonumber \\
&& \quad \times \sum_{\bar{l}} \mathcal{C}_{L' -m_{L'} L m_L}^{\bar{l} -m_{L'} + m_L} \mathcal{C}_{l' -m_{l'} l m_l}^{\bar{l} -m_{l'} + m_l} \nonumber \\
&& \quad \times \sum_{ \bar{L}} \mathcal{C}_{L 0 1 0}^{\bar{L} 0} \mathcal{C}_{L m_L 1 \mu}^{\bar{L} m_L + \mu} \tilde{F}_{L l L' l'}^{\bar{l} \bar{L} } (p,q,p',q'). \label{eq:F_func_factorize}
\end{eqnarray}
For general interactions, the function $\tilde{F}$ depends on multiple quantum
numbers $\bar{L}_i$, hence the function takes formally the  form $\tilde{F}_{L l
L' l'}^{\bar{l} \{ \bar{L}_i \} } (p,q,p',q')$. Using this decomposition we can
first precalculate all sums over the projection  quantum numbers $m$ and $\mu_i$
and prestore the result in a function of the form $A_{\beta \beta'}^{\bar{l} \{
\bar{L}_i \}}$. Then the final matrix element  in $LS$-coupling can be
calculated very efficiently via:
\begin{equation}
\left< p q \beta | V_{123} | p' q' \beta' \right> = \sum_{\bar{l}} \sum_{\{ \bar{L}_i \} } A_{\beta \beta'}^{\bar{l} \{ \bar{L}_i \}} \tilde{F}_{L l L' l'}^{\bar{l} \{ \bar{L}_i \} } (p,q,p',q'),
\label{eq:VLS}
\end{equation}
where values of the quantum numbers $L$, $L'$, $l$ and $l'$ are
speficied by the $LS$-coupling partial wave indices $\beta$ and
$\beta'$ (see Eq.~(\ref{eq:LS_bas})). 

Note that by deriving Eq.~(\ref{eq:VLS}) the original problem of calculating
numerically a 5-dimensional integral for  each matrix element as in
Eq.~(\ref{eq:F_func}) has been reduced to the evaluation of a few discrete sums.
The calculation and prestorage  of the matrix elements of $\tilde{F}_{L l L'
l'}^{\bar{l} \{ \bar{L}_i \} } (p,q,p',q')$ can be performed relatively
efficiently since only three  internal integrals have to be performed
numerically. The exact speedup factor of the present method compared to the
conventional approach \cite{Golak_aPWD,Skibinski_aPWD} depends  on the number of
internal sums over $\mu_i$ and $\bar{L}_i$, i.e. on the specific  form of the
interaction. For example, the matrix elements of the chiral long-range
interactions at N$^2$LO proportional to the couplings $c_1$ and $c_3$ can be
calculated with  speedup factors of greater than 1000. Practically, that means
that it is now possible to calculate the matrix elements of all interaction
terms listed in  Table~\ref{tab:PW_data} up to N$^3$LO on a local computer
cluster for sufficiently large basis sizes for studies of few-nucleon scattering
processes, light and medium mass nuclei as well as nuclear matter. Obviously,
the efficiency of the present method decreases  with each additional internal
sum in Eq.~(\ref{eq:spin_factorize}) and Eq.~(\ref{eq:Y_identity}). However, for
all 3NF topologies up to N$^3$LO except the relativistic corrections, we achieve
speedup factors of typically $\gtrsim 100$.

Even though the present algorithm makes explicit use of the local nature of the
chiral 3NFs, it is also possible to treat polynomial non-local terms. This is of
immediate practical importance since the relativistic corrections at N$^3$LO
have precisely this form~\cite{Bernard_N3LO2}. Consider, for example, a non-local 
momentum structure in the center of mass frame of the type
\begin{equation}
(\mathbf{k}_3 + \mathbf{k}'_3) \cdot (\mathbf{k}_3 - \mathbf{k}'_3) = (\mathbf{q}' + \mathbf{q}) \cdot (\mathbf{q}' - \mathbf{q}).
\end{equation}
Such terms can be treated by factorizing the momentum dependence like 
in Eq.~(\ref{eq:spin_factorize}), for example:
\begin{equation}
\mathbf{q} \cdot \mathbf{q}' = q q' \frac{4 \pi}{3} \sum_{\mu_1,\mu_2 = -1}^1  Y^*_{1 \mu_1} (\hat{\mathbf{\mathbf{q}}}) Y^*_{1 \mu_2} (\hat{\mathbf{\mathbf{q}}}') \, \mathbf{e}_{\mu_1} \cdot \mathbf{e}_{\mu_2}
\end{equation}
and then following exactly the steps like after Eq.~(\ref{eq:spin_factorize}).
Obviously, the algorithm becomes less efficient for non-local interactions, but
the current  framework turns out to be still more efficient than the
conventional approach for the relativistic corrections to chiral 3NF at N$^3$LO.

Generally, 3N interactions can be decomposed in terms of Faddeev components:
\begin{equation}
V_{123} = \sum_{i=1}^3 V_{123}^{(i)}, \label{eq:Faddcomp_def}
\end{equation}
whereas each of the three Faddeev components $V_{123}^{(i)}$ is symmetric in the particle labels $j,k\neq i \in \{1,2,3\}$ and the components are related via permutation transformations:
\begin{equation}
V_{123}^{(2)} = P_{123} V_{123}^{(1)} P_{123}^{-1}, \quad V_{123}^{(3)} = P_{132} V_{123}^{(1)} P_{132}^{-1}.
\end{equation}
Here $P_{123}$ ($P_{132}$) are the permutation operators that permute three
particles cyclically (anti-cyclically) (see Ref.~\cite{Gloeckle_book}). The
decomposition (\ref{eq:Faddcomp_def})  does not uniquely define the Faddeev
components, and the specific values of matrix elements $V_{123}^{(i)}$ are
generally convention-dependent. However, if evaluated between antisymmetrized
wave functions, all  choices and all Faddeev components give the same results.
In contrast, the matrix elements of the completely antisymmetrized interaction,
\begin{equation}
V_{123}^{\text{as}} = ( 1 + P_{123} + P_{132}) V_{123}^{(i)} ( 1 + P^{-1}_{123} + P^{-1}_{132}), \label{eq:V123_as_def}
\end{equation}
are unique. Since the application of the permutation operators in
Eq.~(\ref{eq:V123_as_def}) in a momentum partial wave basis is non-trivial  and
can induce numerical uncertainties (see Ref.~\cite{Skibinski_aPWD}), we
calculate and provide matrix elements of both $\bigl< p q \alpha \bigr|
V_{123}^{(i)} \bigl| p' q' \alpha' \bigr>$ and  $\bigl< p q \alpha \bigr|
V_{123}^{\text{as}} \bigl| p' q' \alpha' \bigr>$. This offers the flexibility to
transform the matrix elements of the Faddeev components to a harmonic
oscillator basis and perform the antisymmetrization directly in this basis. The
resulting matrix elements can then be used for calculations within e.g.~the no-core 
shell model, the valence-shell model, the coupled cluster or the self-consistent Green's
function framework. On the other hand, the antisymmetrized matrix elements
$\bigl< p q \alpha \bigr| V_{123}^{\text{as}} \bigl| p' q' \alpha' \bigr>$ can
be directly used for  studies of infinite nuclear matter or few-body scattering
processes. Due to the large basis sizes shown in  Table~\ref{tab:PW_data}, the
uncertainties induced by the antisymmetrization are very small. We have checked
that for three-nucleon systems the obtained binding energies based on the
Faddeev components and the  antisymmetrized interactions are identical within
the sub-keV level.

For the application of the permutation operator $P_{123}$ in the three-body
plane-wave basis defined in Eq.~(\ref{eq:Jj_bas}) it is important to  implement
this operator in an efficient and numerically stable way. This is in particular
important for calculations in large bases such as those shown in Table
\ref{tab:PW_data}. Several different expressions have been derived for the
permutation operator (see, e.g.,~\cite{Gloeckle_book, Gloeckle_3Ncont}). All
these expressions suffer from problems due to ratios of possibly very small
numerators and denominators, which can, lead to numerical
instabilities for partial waves with large angular momenta. For our
calculations, we use a novel improved implementation. Following the derivations
of Ref.~\cite{Gloeckle_book}, it is straightforward to derive the following
expression:
\begin{eqnarray}
&& \left< p q \alpha | P_{123} | p' q' \alpha' \right> = \sum_{\mathcal{L}, \mathcal{S}} \sqrt{\hat{J} \hat{j} \hat{J}' \hat{j}'} \hat{\mathcal{S}} \nonumber \\
&& \times \left\{
\begin{array}{ccc}
L & S & J \\
l & 1/2 & j \\
\mathcal{L} & \mathcal{S} & \mathcal{J}
\end{array}
\right\}
\left\{
\begin{array}{ccc}
L' & S' & J' \\
l' & 1/2 & j' \\
\mathcal{L} & \mathcal{S} & \mathcal{J}
\end{array}
\right\} \nonumber \\
&& \times (-1)^{S'} \sqrt{\hat{S} \hat{S}'} 
\left\{
\begin{array}{ccc}
1/2 & 1/2 & S \\
1/2 & \mathcal{S} & S'
\end{array}
\right\} \nonumber \\
&& \times (-1)^{T'} \sqrt{\hat{T} \hat{T}'} 
\left\{
\begin{array}{ccc}
1/2 & 1/2 & T \\
1/2 & \mathcal{T} & T'
\end{array}
\right\} \nonumber \\
&& \times 8 \pi^2 \int d \cos \theta_{\mathbf{p} \mathbf{q}} \frac{\delta(p' - |1/2 \mathbf{p} + 3/4 \mathbf{q}|)}{p'^2} \frac{\delta(q' - |\mathbf{p} - 1/2 \mathbf{q}|)}{q'^2} \nonumber \\
&& \times \sum_{m_{\mathcal{L}}}  \mathcal{Y}_{L l}^{\mathcal{L} m_{\mathcal{L}}} (\hat{\mathbf{p}} \hat{\mathbf{q}}) \mathcal{Y}_{L' l'}^{\mathcal{L} m_{\mathcal{L}}} (\widehat{- 1/2 \mathbf{p} - 3/4 \mathbf{q}},\widehat{\mathbf{p} - 1/2 \mathbf{q}}). \label{eq:P123_matrixelements}
\end{eqnarray}
The key difference to other expressions is the fact that we directly perform the
angular integrals as shown in Eq.~(\ref{eq:P123_matrixelements}) without
decomposing the angular dependence of the spherical harmonic functions  any
further. This implementation turns out to be numerically more efficient and,
most importantly, is perfectly stable even for large values of angular momenta.
The matrix elements of the operator $P_{132}$ are also given by
Eq.~(\ref{eq:P123_matrixelements}) (see Ref.~\cite{Gloeckle_book}).

\begin{figure*}[t]
\centering
\includegraphics[scale=0.43]{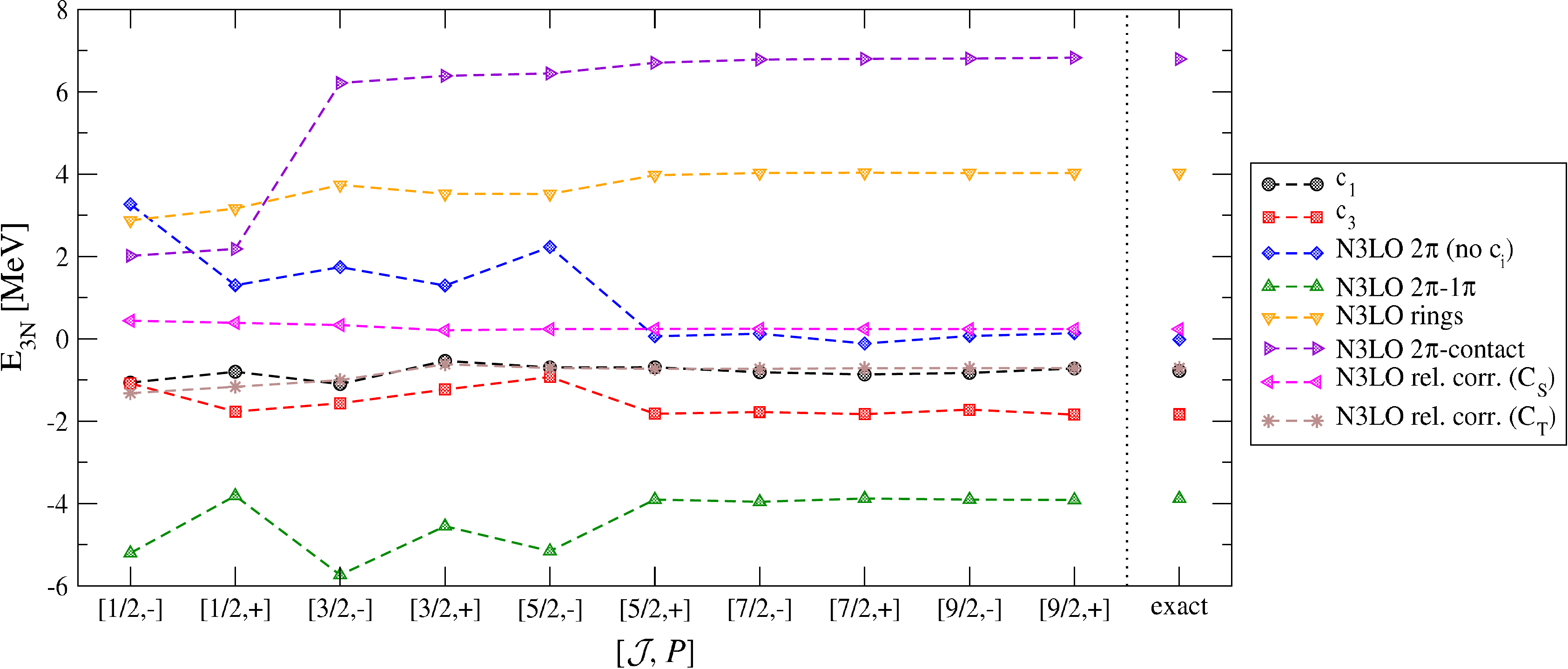}
\caption{(color online) Partial wave contributions to the energy per particle to neutron matter in the Hartree-Fock 
approximation at nuclear saturation density ($k^n_F = 1.7 \, \text{fm}^{-1}$) for the individual 3NF topologies 
up to N$^3$LO. For the shown energies we use the coupling constants $C_S = C_T = 1$ and $c_i = 1 \, \text{GeV}^{-1}$. 
All results show the accumulated energy contributions, including all contributions up to the given partial-wave channel. 
Here $P$ denotes the three-body parity $P = (-1)^{L + l}$ and $\mathcal{J}$ is the three-body total angular momentum, 
as defined in Eq.~(\ref{eq:Jj_bas}). In neutron matter only channels with $\mathcal{T}=3/2$ contribute. The exact 
benchmark results are calculated following Refs.~\cite{fullN3LO,Kruger:2013}.
}
\label{fig:EOS_PNM}
\end{figure*}

\begin{figure*}[t]
\centering
\includegraphics[scale=0.43]{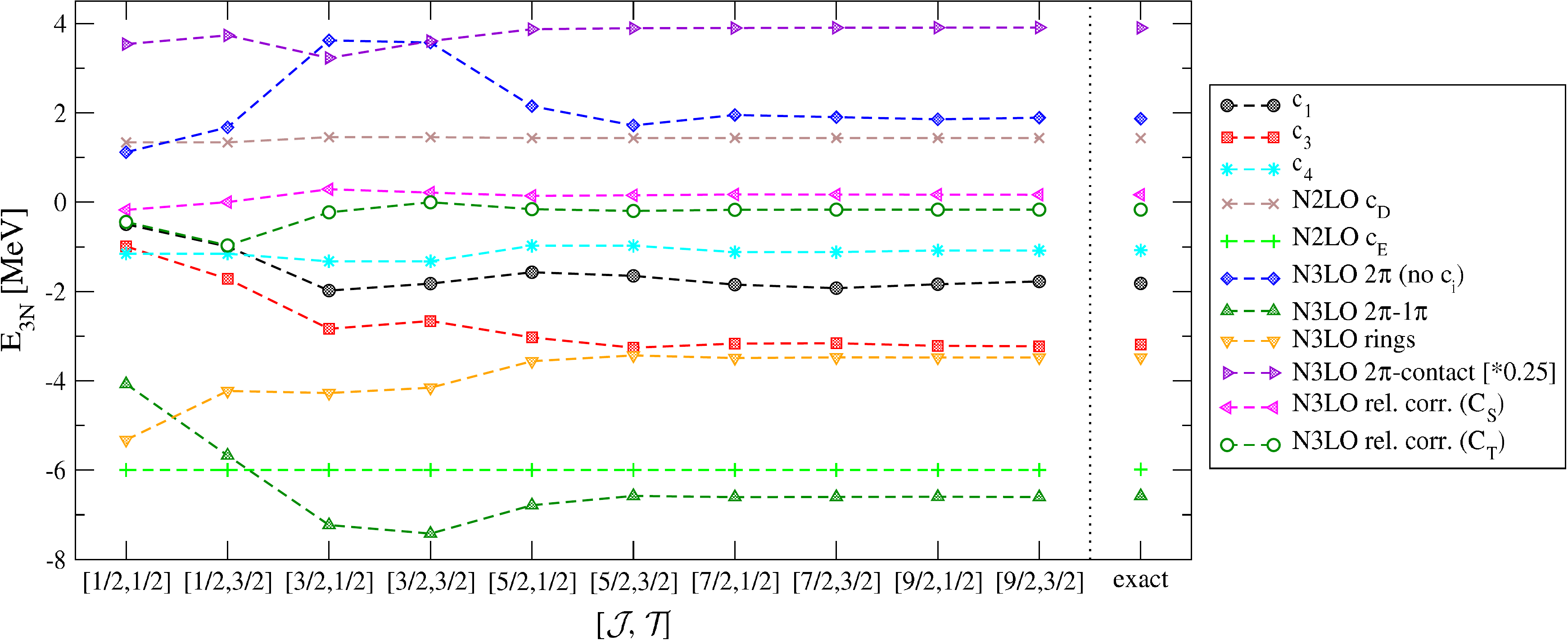}

\caption{(color online) Partial wave contributions to the energy per particle to symmetric 
nuclear matter in Hartree-Fock approximation at nuclear saturation density ($k^n_F = k_F^p = 1.35 \, \text{fm}^{-1}$) 
for the individual 3NF topologies up to N$^3$LO. For the shown energies we use the coupling constants 
$C_S = C_T = 1$ and $c_i = 1 \, \text{GeV}^{-1}$. The results for the $2\pi$-contact topology are scaled 
by a factor of $1/4$ for presentation purposes. All results show accumulated energies, including all 
contributions up to the given partial-wave channel including both three-body parities. $\mathcal{J}$ 
and $\mathcal{T}$ are the three-body total angular momentum and isospin, as defined in Eq.~(\ref{eq:Jj_bas}). 
The exact benchmark results are calculated following Ref.~\cite{Kruger:2013}.}
\label{fig:EOS_SNM}
\end{figure*}

As a first application, we use the new framework to calculate the matrix
elements of the chiral 3NF up to N$^3$LO.  The individual topologies are shown
in Fig.~\ref{fig:topologies}. In addition to the pion decay coupling $F_\pi$
and nucleon axial-vector coupling constant $g_A$, the chiral 3NFs up to
N$^3$LO  depend on 7 LECs in total, namely $c_1, c_3, c_4, c_D, c_E, C_S$ and
$C_T$ \cite{Ishikawa:2007zz,Bernard_N3LO1,Bernard_N3LO2,Epelbaum:2002vt}, for
which we  want to keep the flexibility of being able to change their values
after performing  the partial-wave decomposition. Specifically, the individual
topologies are\footnote{We correct  a misprint in Eq.~(4.14) of
Ref.~\cite{Bernard_N3LO2} and use the parameter values  $\beta_8=\frac{1}{4}$
and $\beta_9 = - \frac{1}{4}$, corresponding to the  "minimal nonlocality"
choice of the potential (see Ref.~\cite{improvedlocalNN}).} (see also
Fig.~\ref{fig:topologies}): (a) 2$\pi$ exchange [$c_1, c_3, c_4, 1$], (b)
1$\pi$-contact [$c_D$], (c) pure contact [$c_E$], (d) 2$\pi$-1$\pi$ exchange
[1], (e) ring contributions $[1]$, (f) 2$\pi$-contact [$C_T$] and (g)
relativistic corrections [$C_S, C_T, 1$]. Here, the couplings in square
brackets denote those free LECs which appear in each of these topologies,
whereas the constant '$1$' indicates contributions that do not contain any of
the indicated 7 LECs. This leads in total to 12 individual contributions for
each three-body partial wave as defined in Eq.~(\ref{eq:Jj_bas}). 

For each of these contributions we first compute one of the Faddeev 
components $\bigl< p q \alpha \bigr| V_{123}^{(i)} \bigl| p' q' \alpha' \bigr>$ for all partial wave
channels and basis sizes shown in Table \ref{tab:PW_data}.  Subsequently, for
the calculation of the antisymmetrized matrix elements  $\bigl< p q \alpha
\bigr| V_{123}^{\text{as}} \bigl| p' q' \alpha' \bigr>$ we apply the
permutation operator as defined in Eq.~(\ref{eq:P123_matrixelements}). The
application of the permutation operator in the partial wave basis involves, in
principle, a complete sum over intermediate partial wave quantum numbers (see,
e.g., Ref.~\cite{Skibinski_aPWD}). Thanks to the large basis sizes shown in
Table~\ref{tab:PW_data},  we ensure that the antisymmetrization leads to 
well-converged results. Furthermore, it is straightforward to  generate arbitrary
other products of the permutation operator $P_{123}$ and the Faddeev
components in a computationally very efficient way. Such products appear, for
example, in Faddeev equations for few-body scattering
problems~\cite{Gloeckle_3Ncont}.

\section{Application of chiral 3N forces at N$^3$LO to nuclear matter and $^3$H}
\label{sec:app_3NF}

\begin{figure*}[t]
\centering
\includegraphics[scale=0.55]{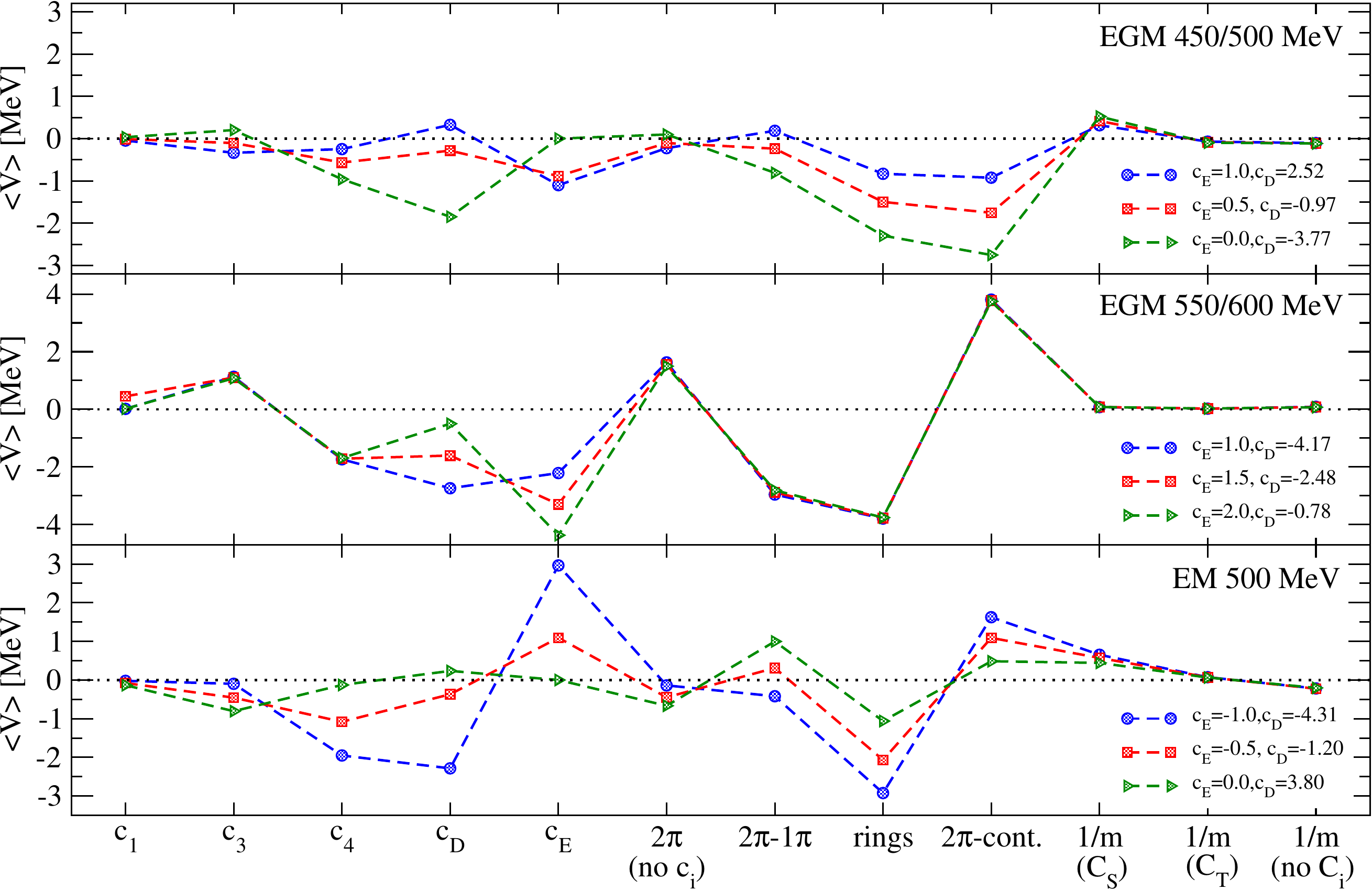}
\caption{(color online) Contributions of the individual topologies to the triton energy for three different 
NN interactions of Refs.~\cite{Progpart_Epelbaum,EM}. The LECs $c_D$ and $c_E$ are chosen to be of natural size 
and are fitted to the experimental triton binding energy. The topologies $c_1$, $c_3$ and $c_4$ contain 
contributions from N$^2$LO and N$^3$LO, $c_D$ and $c_E$ are pure N$^2$LO contributions and 
$2\pi (\text{no } c_i)$, $2\pi$-$1\pi$, rings, $2\pi$-cont and $1/m$ corrections are N$^3$LO contributions. See main text for details.}
\label{fig:3H_topologies}
\end{figure*}
Figs.~\ref{fig:EOS_PNM} and \ref{fig:EOS_SNM} illustrate the convergence of the
partial wave expansion of the calculated chiral 3NFs. These figures show the 3NF
contributions to the energy per particle of neutron matter (Fig. 2) and
symmetric nuclear matter (Fig. 3) in Hartree-Fock approximation for the
individual partial-wave channels and 3NF topologies. These results provide
direct insight in the required  number of partial-wave channels in mean-field
calculations. Of course, the number of required partial-wave channels has to be
checked in more  detail in realistic nuclear structure calculations, for which
also many-body correlations are important. Moreover, these results also serve as
a direct benchmark of the calculated matrix  elements, since the energy
contributions in the Hartree-Fock approximation have been already calculated
independently based on the operator expressions directly, see
Refs.~\cite{fullN3LO,Kruger:2013} for details. 

Specifically, for the results shown in Fig.~\ref{fig:EOS_PNM} we use $k_{\rm{F}}=1.7 \, \text{fm}^{-1}$ for
the neutron Fermi momentum, which corresponds to a neutron number density of $n
\simeq 0.166 \, \text{fm}^{-3}$.  Since the Fermi momentum serves as an
ultraviolet cutoff, we do not need to regularize the interactions for these
benchmark calculations. In neutron matter only matrix elements in the three-body
isospin channel $\mathcal{T} = 3/2$ contribute, whereas the N$^2$LO topologies
that include the low-energy coupling constants $c_4$, $c_D$ and $c_E$ vanish
exactly~\cite{HebelerPNM}. The detailed expressions for the Hartree-Fock
energies for neutron matter in terms of the antisymmetrized matrix elements
$V_{123}^{\text{as}}$ are given in Ref.~\cite{Hebeler_PNMevolved}. The results
of Fig.~\ref{fig:EOS_PNM} show that matrix elements up to the partial wave
channel with $\mathcal{J} = 5/2$ and both three-body parities $P=(-1)^{L+l}$ can
provide  significant contributions to the energy, whereas higher partial waves
give only small corrections. Overall, we find that the results including all
contributions up to $\mathcal{J}=9/2$ are very well converged and show
excellent agreement with the exact results (see also~\cite{Christian_masterthesis}).

For symmetric nuclear matter we find a very similar convergence pattern. In
contrast to neutron matter here all 3NF topologies shown in
Fig.~\ref{fig:topologies} and also both three-body isospin channels,
$\mathcal{T}=1/2$ and $\mathcal{T}=3/2$, contribute to the energy. For the
results shown in Fig.~\ref{fig:EOS_SNM} we fix the neutron and proton Fermi
momenta to $k^n_F = k_F^p = 1.35 \, \text{fm}^{-1}$,  which again corresponds to
a total number density of $n \simeq 0.166 \, \text{fm}^{-3}$. We show the
contributions to the energy for the individual partial-wave channels, whereas
here  each $[\mathcal{J},\mathcal{T}]$ channel includes contributions from both
three-body parity channels $P=\pm 1$. Again, we observe excellent partial wave
convergence and essentially perfect agreement with the exact Hartree-Fock
results.

Next, in Fig.~\ref{fig:3H_topologies} we illustrate the contributions of the
individual topologies to the binding energy of $^3$H. In order to probe  scheme
dependence  of our results, the present calculations use three different NN
interactions: the N$^3$LO potentials of Ref.~\cite{Epelbaum:2004fk} with the
cutoff combinations  $\Lambda/\tilde{\Lambda} = 450/500 \text{MeV}$ and $550/600
\text{MeV}$ (EGM) and the N$^3$LO potential of Ref.~\cite{EM} (EM).  For our
calculations we fix the values of  the LECs $c_1, c_3, c_4, C_S$ and $C_T$
consistently to their values of the NN interactions. Notice further that the
N$^3$LO corrections to the two-pion exchange topology involve contributions
which account for finite shifts in the LECs $c_i$ in the N$^2$LO 3NF
expressions~\cite{Bernard_N3LO1}. All these effects are properly taken into
account.  The couplings $c_D$ and $c_E$ are fixed to the experimental binding
energy of $^3$H and by requiring the values of  those LECs to be of natural
size. For each of the NN potentials we used three sets of values, their specific
values for the different interactions are given in  the legend of
Fig.~\ref{fig:3H_topologies}. The matrix elements of the 3NF are regularized by
applying a non-local regulator of the form $f_R(p,q) = \exp[-((p^2 + 3/4
q^2)/\Lambda_{\rm 3N}^2)^3]$  as in Ref.~\cite{Epelbaum3N}, whereas the value of the
cutoff $\Lambda_{\rm 3N}$ is chosen to be consistent with the cutoff value $\Lambda$
of the corresponding NN interaction.  Note that all results of this figure are based on
the consistent wave functions for each of these fits, which obviously
complicates a detailed quantitative comparison  of results based on different
fits or NN interactions. However, we do not expect that this affects the main
qualitative features of the results, which can be summarized as follows: first,
the size of the  contributions of the individual topologies depends sensitively
on the employed NN interaction as well as on the values of the LECs $c_D$ and
$c_E$. Second, the contributions of topologies at N$^3$LO are not
suppressed compared to those at N$^2$LO for the presently employed 3N
interactions. Similar findings have been found before for $^3$H using a
restricted number of partial waves \cite{Skibinski:2011vi} and  for the mean-field 
contributions of the N$^3$LO 3NFs  to neutron matter and symmetric
matter~\cite{fullN3LO,Kruger:2013}. Third, the perturbativeness of the 3NFs also
depends sensitively on the employed NN interaction. While, e.g.,  for the
interaction EGM 550/600 the wave function is not strongly affected by variations
of the short-range couplings, for the other two NN interactions the
contributions from the long-range 3N topologies also change when the LECs
$c_D$ and $c_E$ are varied.

When interpreting these results, it is important to keep in mind that neither
the individual nor the total contribution of the 3NFs to the binding energies of
nuclei are experimentally observable. The considered contributions of individual
topologies represent, strictly speaking, bare quantities while all estimations
based on power counting refer to renormalized ones, see also
Ref.~\cite{Skibinski:2011vi}. Moreover, the employed nonlocal regulators do
not actually cut off all short-range pieces of the 3NFs, so even the long-range 
3NF topologies do contain short-range contributions after regularization
(see Ref.~\cite{improvedlocalNN} for a related discussion). 

Clearly, expectation values of such short-range admixtures are  strongly scheme
dependent. Recently, a novel type of chiral NN interactions has been
developed~\cite{localMCPRL,localMClong,improvedlocalNN}. For these new
interactions, the long-range topologies are regularized by local regulators that
act only on the relative particle distance in coordinate space and allow for a
cleaner separation  of long- and short-distance physics. Indeed, in
Ref.~\cite{Epelbaum:2014sea} it was shown explicitly that 2$\pi$-exchange
contributions dominate at long- and intermediate distances when local
regulators are employed for NN interactions.

\section{Summary and outlook} 
\label{sec:summary}

In summary, the new framework presented in this work makes it possible to
include the chiral 3NF at N$^3$LO and beyond in ab initio few- and many-body
calculations. For non-local regulators, the calculated matrix elements can be
immediately used within various many-body frameworks for novel studies including
all NN and 3N contributions consistently beyond N$^2$LO. The application of the
regularization scheme employed in Ref.~\cite{improvedlocalNN, N4LO_NN} for the NN
forces to 3NF will require a generalization of the present framework. In
contrast to the considered non-local regulators, which only depend on the
magnitude of the Jacobi momenta and act just as multiplicative factors in
momentum space, local regulators do mix different partial waves. The application
of these regulators in momentum space can, e.g., be implemented using their
partial wave decomposed form and performing numerical evaluation of the
corresponding folding integrals. Work along this line is in progress.

We also emphasize that a careful investigation of systematic uncertainties in few- and
many-nucleon calculations represents an important goal in nuclear physics, see
also discussions in Refs.~\cite{improvedlocalNN,Dickuncertainties}. The
availability of matrix elements of NN and 3N interactions   at different orders
in the chiral expansion and within different regularization schemes will provide
an important contribution towards a better understanding of the systematic
uncertainties and ultimately allow for precision tests of chiral dynamics in
nuclear systems. Last but not least, we emphasize that the presented framework
can be straightforwardly applied to 3NF at order N$^4$LO~\cite{N4LO1,N4LO2} and
to EFT interactions with explicit $\Delta$-degrees of freedom
\cite{Epelbaum:2007sq}.

\begin{acknowledgments} We thank V.\ Durant, R.\ J.\ Furnstahl, A.\ Schwenk for helpful comments, 
T.\ Kr\"uger and I.\ Tews for providing the exact benchmark Hartree-Fock results for
neutron matter and nuclear matter, K.\ Topolnicki and H.\ Wita{\l}a for their 
contribution to the partial-wave framework we used for testing the calculated
matrix elements, and C.\ Drischler for performing additional
test calculations. The present work was supported by the ERC Grants No. 
307986 STRONGINT and 259218 NUCLEAREFT, by the Helmholtz Alliance HA216/EMMI, 
by the European Community-Research Infrastructure Integrating 
Activity ``Study of Strongly Interacting Matter'' (acronym HadronPhysics3, 
Grant Agreement n. 283286) under the Seventh Framework Programme of EU, and by 
awards of computational resources from the Ohio Supercomputer Center and from 
the J\"ulich Supercomputing Center. The work of the Cracow group was supported 
by the Polish National Science Center under Grant No.DEC-2013/10/M/ST2/00420.

\end{acknowledgments}

\bigskip

\appendix
\def\theequation{A\arabic{equation}}
\setcounter{equation}{0}
\section{Appendix:\\Angular integrations in partial wave decomposition}

\label{pwd_derivation}
In this Appendix we describe the integral transformations which allow
for a decoupling of the three non-trivial integrations over spherical
harmonics in the partial wave decomposition of the 3NF from the other
five which can be performed analytically. 
Let us start with the Eq.~(\ref{eq:F_func}) and add a radial integration 
over $\tilde{p}^\prime$ and $\tilde{q}^\prime$ in order to have a translationally invariant measure. 
We can achieve this by introducing additional integrations with delta-functions via

\begin{eqnarray}
&& F_{L l L' l'}^{m_L m_l m_{L'} m_{l'}} (p, q, p', q') 
= \frac{1}{p^{\prime\,2} q^{\prime\,2}}\int d^3 \tilde{q}^\prime d^3 \tilde{p}^\prime d\hat{\mathbf{q}} d\hat{\mathbf{p}}
\nonumber\\
&& \quad \times \delta(p^\prime - \tilde{p}^\prime\,) \delta(q^\prime - \tilde{q}^\prime\,) Y_{L' m_{L'}}^{*} (\hat{\tilde{\mathbf{p}}}^\prime\,) Y_{l' m_{l'}}^{*} (\hat{\tilde{\mathbf{q}}}^\prime\,) Y_{L m_L} (\hat{\mathbf{p}}) Y_{l m_l} (\hat{\mathbf{q}}) \nonumber\\
&&\quad \times V_{123}^{\rm{loc}} (\tilde{\mathbf{p}}^\prime - \mathbf{p}, \tilde{\mathbf{q}}^\prime - \mathbf{q}). 
\end{eqnarray}
Now we can make a substitution
\begin{equation}
\tilde{\mathbf{p}}^\prime \rightarrow \tilde{\mathbf{p}}^\prime + \mathbf{p} \quad {\rm and}\quad \tilde{\mathbf{q}}^\prime \rightarrow \tilde{\mathbf{q}}^\prime + \mathbf{q}, 
\end{equation}
which leads to 
\begin{eqnarray}
&& F_{L l L' l'}^{m_L m_l m_{L'} m_{l'}} (p, q, p', q') 
= \frac{1}{p^{\prime\,2} q^{\prime\,2}}\int d^3 \tilde{q}^\prime d^3 \tilde{p}^\prime d\hat{\mathbf{q}} d\hat{\mathbf{p}}
\nonumber\\
&& \quad \times \delta(p^\prime - |\tilde{\mathbf{p}}^\prime+\mathbf{p}|\,) \delta(q^\prime - |\tilde{\mathbf{q}}^\prime + \mathbf{q}|\,)
V_{123}^{\rm{loc}} (\tilde{q}^\prime,\tilde{p}^\prime,\cos \theta_{\tilde{\mathbf{p}}^\prime \tilde{\mathbf{q}}}^\prime)\nonumber\\
&& \quad \times Y_{L' m_{L'}}^{*} (\widehat{\tilde{\mathbf{p}}^\prime+\mathbf{p}}\,) Y_{l' m_{l'}}^{*} (\widehat{\tilde{\mathbf{q}}^\prime + \mathbf{q}}\,) Y_{L m_L} (\hat{\mathbf{p}}) Y_{l m_l} (\hat{\mathbf{q}}). 
\end{eqnarray}
In this form it is manifest that the integrals factorize and we need to calculate
\begin{eqnarray}
&& \frac{1}{p^{\prime\,2} q^{\prime\,2}}\int d\hat{\mathbf{q}} d\hat{\mathbf{p}} d\hat{\mathbf{p}}^\prime \int_0^{2\pi} d\phi_{\tilde{q}^\prime} \delta(p^\prime - |\tilde{\mathbf{p}}^\prime+\mathbf{p}|\,) \delta(q^\prime - |\tilde{\mathbf{q}}^\prime + \mathbf{q}|\,)\nonumber\\
&& \quad \times Y_{L' m_{L'}}^{*} (\widehat{\tilde{\mathbf{p}}^\prime+\mathbf{p}}\,) Y_{l' m_{l'}}^{*} (\widehat{\tilde{\mathbf{q}}^\prime + \mathbf{q}}\,) Y_{L m_L} (\hat{\mathbf{p}}) Y_{l m_l} (\hat{\mathbf{q}}),
\end{eqnarray}
where $\phi_{\tilde{q}^\prime}$ is an azimutal angle of $\hat{\tilde{\mathbf{q}}}^\prime$. This integral does not involve the 3NF and can be calculated analytically. The full result is given by Eq.~(\ref{final_result_pwd}).

\end{document}